# Creative Gardens

Gerard BRISCOE and Joseph LOCKWOOD

Queen Mary University London, Glasgow School of Art

Can we move beyond simply networking creative individuals to establishing diverse *communities of practice* for innovation through discursive methods. Furthermore, can we digitise their creativity activities within an integrative socio-cultural collaborative technology platform that could then support distributed innovation. First, we consider the complexity of creative cultures from the perspective of design innovation, including how to nurture creativity activities in what we call Creative Gardens. Specifically, how they could grow, diverge, and combine, being cultivated to nurture emergent, disruptive, collaborative innovation. Then, we consider the digitisation of Creative Gardens from the perspective of digital culture. Specifically, the tenets of Creative Gardens as dynamic and innovative communities. This includes considering the challenges and opportunities around digitisation, the influences around the connectivity with knowledge cultivation, and the potential for distributed innovation as collective intelligence to utilise diverse expertise. We conclude be considering the importance of the issues and questions raised, and their potential for the future.

## Introduction

The growing importance and profile of innovation theory and policy has also evolved in line with developing thinking about the scope and nature of innovation in a modern economy. The central role of innovation in creating future prosperity and quality of life is widely acknowledged and accepted (OECD 2010). However, innovation has multiple-drivers, social, cultural, organisational as well as technological. Furthermore, the linear model of innovation through science, 'research and development', and technology development has been augmented through the exploration of open innovation models, the importance of the creative economy and interdisciplinary approaches. Innovation often comes from looking sideways, to seek ideas in adjacent fields or disciplines, which when abducted into your own domain might yield new insight or combination. This process of combination often relies on people who span different cultures and disciplines and spaces where ideas and people mingle, in which creativity comes from interaction and dialogue between different ideas not just from diversity alone (Leadbeater 2006).

Within 'cultures of innovation', we seek to investigate approaches to create capacity for sustainable innovation. We consider 'a culture of innovation' as a complex adaptive system that have a large numbers of component that interact and adapt or learn (Holland 2006), because it has been suggested (Griggin & Stacey 2005) 'innovation is pursued as the novelty that emerges from conversations collaborations in dynamic, non-linear, networked communities'. Also, we need to understand the social goals like creativity, personal satisfaction and freedom in how we begin to reorganise for innovation. We must also be mindful of the social fabric of the organisation and recognise culture as a powerful and sophisticated agent (Alvesson 2012). Furthermore, to unlock creative potential we should take a situational rather than a dispositional view of leadership to enable a field of 'creative leadership, by igniting the collective creativity from the bottom up' (Radjou, Prabhu, Kaipa & Ahuja 2010). Therefore, we need to rethink how we organise work for innovation is led by a technological disruption however; technology only enables change if wedded to people's need and desires.

We propose Creative Gardens to explore achieving innovation through creative activities. Our model advocates a holistic hybrid approach to nurturing creativity for innovation, ultimately leading to socio-cultural capital capable of addressing crowdsourced areas of interest. This curation approach to disruptive innovation would integrate open distributed innovation with communities of practice. Forms of cultivation would include encouraging design innovation, cluster building, diverse grouping, and disruptive collaboration. These would ultimately aim to stimulate the growth of the socio-cultural capital needed to facilitate the creative capability required to achieve innovation on range of applications. For example, in the context of sustainability Creative Gardens would stimulate the growth of the socio-cultural capital needed to facilitate the creative capability required to achieve the sustainable use of natural resources primarily through replenishment, as preservation alone would be insufficient. Specifically, contributing to innovation aimed at cultivating growth in new socio-cultural capabilities (capital) that replenish natural capital, rather than economic growth that is destructive of natural capital. Digitising Creative Gardens requires considering the physical and digital interaction that would be utilised by designers to achieve an intuitive community engagement experience in the complexity of creative spaces. This will include how they could form, operate, diverge, and merge to facilitate cultures of creativity for community-driven emergent disruptive innovation. Therefore, we aim to develop the insight necessary for a conceptual design allowing for the loss-less digitisation of Creative Gardens to support distributed communities of practices in achieving innovation. For example, in promoting networks-of-networks that favour collective creativity, alongside multidisciplinary expertise that is spatially and temporally located. In other words, a distributed approach consisting of place-based resources, which are part of a design innovation solution.

**Cultures of Innovation**

Cultures of innovation require a discussion on complexity and innovation, because innovation can be considered as being on the *edge of chaos*. This *edge of chaos* in complexity is most frequently associated with work concerned with living systems (e.g. insect colonies, the human body, neural networks, etc). Complex, non-linear dynamic systems with rich networks of interacting elements have a zone of operation (states) that lies between chaotic and near static behaviour ones with minimal spontaneous activity (Goodwin 2007). Such systems on the *edge of chaos* appear constantly to adapt, self-organising to create configurations that ensure compatibility with an ever-changing environment. So, humanity is now evolving from the hierarchal structure of industrial culture to a network structure of robust, creative and locally empowered societies (Goodwin 2007). This implies living in social networks with non-hierarchical connectedness, with maximum freedom to the individual and maximum potential for the collective.

Humanity spent the last three hundred years discovering the *real truth*, i.e. the laws understood by mechanical causality, resulting in prediction and control. This was reliable for forming knowledge on technical process, allowing the scales of industrialisation. However, this approach is less helpful when more holistic, creative and unpredictable approaches are required to manage situations and solve problems (e.g. sustainability). Also, while a strictly quantitative approach to nature has given us the ability to produce vast quantities of consumer goods and wealth, it has resulted in the destruction of species and people globally (Goodwin 2007). So, we continue to attempt to understand the post-industrial emerging network structure of society with established mechanistic rational approaches to innovation. Therefore, a latent need is developing to focus on enabling concepts and methods that would allow us to address the dichotomy this increasingly presents between current approaches to innovation and cultural needs.

It has been suggested that innovation is pursued as the novelty that emerges from collaborations in these dynamic, non-linear and networked communities (Griggin & Stacey 2005). So, in this network structure creativity is increasingly a form of agency (Nussbaum 2011), and therefore has the potential to drive cultural evolution (Csikszentmihalyi 1998). Therefore, we need to experiment with extreme collaboration for networked collectives, challenging traditional models of single disciplinarily and avoiding *silos* of knowledge. Furthermore, we need to consider how digital technologies could facilitate *distributed* collective creatives that we call Creative Gardens. We could rely on self-organisation within social networks to spawn the collective creatives of Creative Gardens.

## Digital Cultures

A digital platform for Creative Gardens requires considering the integration of digital technologies with Creative Gardens, and the digital cultures affecting their distributed innovation and creativity. Continuous analogue activities that are digitised to their discrete digital counterparts often presents challenges, interacting with digital cultures in new and unexpected ways. So, digitisation can also lead to the creation of new aspects to digital cultures. Digitisation of any cultural artefacts results in the loss of context (or aura) in which it originally existed (Frascina 1992). Digital reproduction has created the potential for any cultural artefact to be seen anywhere, which inevitably has an effect on the meaning (Manovich 2001). For example, digital technologies of reproduction (e.g. photography, film, audio recordings, etc) uproot cultural artefacts form their situational contexts. Distance, space, perspective, horizon, history, geography, etc, all create context (Miller 2011). This is especially a risk in digitising creative activities, because the critical elements are not themselves well understood and so may be easily lost. So, the challenge with Creative Gardens is to understand and appreciate the critical aspects of context that could be lost. Then, either to recreate them or compensate for their loss. Alternatively, we could ensure that use occurs with an appropriate offline and on-line balance, for example digitising the results of creative activities for distribution, rather than the activities themselves.

The current emphasis on allowing knowledge cultivation to be strictly owned as a source of profit generally results in the established owners of capital, rather than creators of knowledge, benefitting from intellectual property rewards (Miller 2011). This Information Feudalism (Drahos & Braithwaite 2002) represents the concern that the current emphasis on strong intellectual property rights leads to a situation where the cultivation and ownership of information becomes increasingly concentrated (Briscoe & Marinos 2009, Stanley & Briscoe 2010, Marinos & Briscoe 2009). The risk for Creative Gardens is that digitisation leads to the adoption a digital culture's norms regarding knowledge ownership, and that is inappropriate to the decentralised open knowledge cultivation approach before digitisation. So, if a model of open source software were adopted we would consider the digitisation to be *loss-less* with regards to knowledge cultivation. However, if an approach akin to software patents were adopted we would consider it to be critically *lossly*.

Another significant aspect of digital culture for Creative Gardens is *collective intelligence* (Miller 2011), which is the spontaneous self-organisation of expertise, resources and information through digital technologies towards collective creation for problem solving (Rheingold 2002, Leadbeater 2010, Herz 2005). It is a new form of knowledge cultivation arising from the interactivity and networking of convergence media (e.g. wikis and wikipedia). Participants can *pool* their interests and expertise towards the solving of shared problems, creating a *digital commons* for mutual benefit. An example of this is crowdsourcing, which could be used in forming Creative Gardens around shared areas of interest, creating a Digital Ecosystem of Creative Gardens (Briscoe, Sadedin & De Wilde 2011, Briscoe 2009, Briscoe & De Wilde 2006). A Digital Ecosystem is a distributed, adaptive, open socio-technical system with properties of self-organisation, scalability and sustainability inspired from natural ecosystems. Participants could then nurture their Creative Gardens to encourage emergent disruptive innovation through design innovation, cluster building and diverse grouping. Furthermore, participants could diverge their Creative Gardens into two or more, for example to better support competing ideas or approaches to innovation. Also, merging two or more of their Creative Gardens, if for example they discover they possess innovations capable of enhancing one another. While this role of curation would be fulfilled by the participants, it could also be augmented by platform recommendations like (Briscoe & De Wilde 2008). For example, the system could recommend merging Creative Gardens based upon them sharing a similar focus.

## Conclusion

We have attempted to provide a sufficiently complete discussion of balancing the social, cultural, and technological concerns affecting the design and architecture of Creative Gardens. We built upon understanding a mixture of epistemologies that are concerned with innovation, specifically design and digital innovation. We considered digitised Creative Gardens that would be formed through crowdsourcing, and curated to be merged, diverged and grown. Furthermore, we believe that the research questions we have raised here are in themselves an interesting contribution to the development of digital technologies for innovation and creativity. We consider that the Internet of today will come to be viewed as a primitive data network, when the questions we have raised have been answered and paradigms like Creative Gardens become established.